\documentclass[aps,prl,superscriptaddress,floatfix,twocolumn]{revtex4}
\usepackage{bm}
\usepackage{graphicx}
\usepackage{amsfonts}
\usepackage{amsmath}
\usepackage{amssymb}

\begin{document}

 \title{Reentrant behavior of the phase stiffness in Josephson
junction arrays}

\author{Luca Capriotti}
\affiliation{Kavli Institute for Theoretical Physics, University
    of California, Santa Barbara, CA 93106, USA}
\author{Alessandro Cuccoli}
\affiliation{Dipartimento di Fisica dell'Universit\`a di Firenze -
             via G. Sansone 1, I-50019 Sesto Fiorentino (FI), Italy}
\affiliation{Istituto Nazionale per la Fisica della Materia
             \,-\,U.d.R. Firenze -
             via G. Sansone 1, I-50019 Sesto Fiorentino (FI), Italy}
\author{Andrea Fubini}
\affiliation{Dipartimento di Fisica dell'Universit\`a di Firenze -
             via G. Sansone 1, I-50019 Sesto Fiorentino (FI), Italy}
\affiliation{Istituto Nazionale per la Fisica della Materia
             \,-\,U.d.R. Firenze -
             via G. Sansone 1, I-50019 Sesto Fiorentino (FI), Italy}
\author{Valerio Tognetti}
\affiliation{Dipartimento di Fisica dell'Universit\`a di Firenze -
             via G. Sansone 1, I-50019 Sesto Fiorentino (FI), Italy}
\affiliation{Istituto Nazionale per la Fisica della Materia
             \,-\,U.d.R. Firenze -
             via G. Sansone 1, I-50019 Sesto Fiorentino (FI), Italy}
\author{Ruggero Vaia}
\affiliation{Istituto di Fisica Applicata `Nello Carrara'
             del Consiglio Nazionale delle Ricerche,\\
             via Madonna del Piano, I-50019 Sesto Fiorentino (FI), Italy}
\affiliation{Istituto Nazionale per la Fisica della Materia
             \,-\,U.d.R. Firenze -
             via G. Sansone 1, I-50019 Sesto Fiorentino (FI), Italy}
\date{\today}

\begin{abstract}
The phase diagram of a 2D Josephson junction array with large
substrate resistance, described by a quantum $XY$ model, is
studied by means of Fourier path-integral Monte Carlo. A genuine
Berezinskii-Kosterlitz-Thouless transition is found up to a
threshold value $g^\star$ of the quantum coupling, beyond which no
phase coherence is established. Slightly below $g^\star$ the phase
stiffness shows a reentrant behavior with temperature, in
connection with a low-temperature disappearance of the
superconducting phase, driven by strong nonlinear quantum
fluctuations.
\end{abstract}


\maketitle

Two-dimensional Josephson junction arrays (JJA) are among the
best experimental realizations of a model belonging to the $XY$
universality class and offer the possibility of controlling and
studying a variety of phenomena related to both the dynamics and
the thermodynamics of vortices. In these systems a
Berezinskii-Kosterlitz-Thouless (BKT) transition~\cite{BKT}
separates the superconducting~(SC) and the normal~(N) state, the
latter displaying no phase coherence~\cite{fv01}. For a nanoscale
size of the junctions a new interesting feature
shows up in the JJA, namely the quantum fluctuations of the
superconducting phases. These are caused by the non-negligible
energy cost of charge transfer between SC islands, a consequence
of the small capacitances involved and the fact that phase and
charge are canonically conjugated variables. A relevant effect is
the progressive reduction of the SC-N transition temperature.
Recently, fabricated arrays of nanosized junctions, both
unshunted~\cite{zegm96} and shunted~\cite{yama}, have given the
opportunity to experimentally approach the quantum (zero
temperature) phase transition.

However the mechanism of suppression of the BKT in the
neighborhood of the quantum critical point and its connection with
the observed reentrance of the array resistance as function of the
temperature is not yet clear~\cite{fv01,zegm96,jhog89}. In this
letter we study the SC-N phase diagram by means of path-integral
Monte Carlo (PIMC)~\cite{ccftv02} simulations focusing the
attention on the region of strong quantum fluctuations, in order
to investigate their role in suppressing the BKT transition.

We describe the JJA on the square lattice by a quantum $XY$ model
with the following action
\begin{equation}
 S[\bm\varphi]= \int\limits_0^{\hbar\beta}\!\! du\,
 \bigg\{\!\sum_{\bm{ij}} \frac{\hbar^2C_{\bm{ij}}}{8e^2}
 {\dot \varphi_{\bm{i}}(u)\,\dot\varphi_{\bm{j}}(u)}
  - E_{_{\rm J}}\sum_{\langle{\bm{ij}}\rangle}
  \cos\varphi_{\bm{ij}}(u) \bigg\},
 \label{e.JJA}
 \end{equation}
where $\varphi_{\bm{ij}}=\varphi_{\bm{i}}-\varphi_{\bm{j}}$ is the
phase difference between the $\bm{i}$th and the $\bm{j}$th
neighboring superconducting islands. We assume the presence of
weak Ohmic dissipation due to small currents flowing to the
substrate or through shunt resistances~\cite{yama}, which reflects
into the prescription to consider the phase as an extended
variable~\cite{fv01}. The capacitance matrix reads
$C_{\bm{ij}}=C\,\big[\eta\,\delta_{\bm{ij}}+(z\,\delta_{\bm{ij}}
-{\textstyle\sum_{\bm{d}}}\delta_{\bm{i},\bm{j+d}})\big]$, where
$C_0\equiv\eta\,C$~\cite{eta} and $C$ are, respectively, the self-
and mutual capacitances of the islands, and $\bm{d}$ runs over the
vector displacements of the $z\,{=}\,4$ nearest-neighbors.

\begin{figure}[t]
\includegraphics[width=80mm]{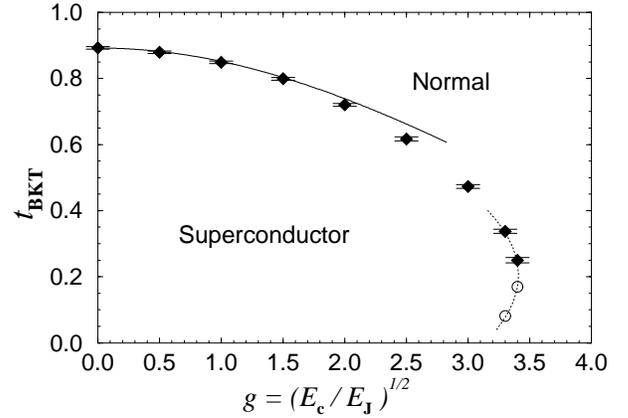}
\caption{Phase diagram for a JJA on a square lattice in the
limit \protect$R_{{\rm{S}}}\gg R_{\rm Q}=h/(2e)^2$ with
$\eta=10^{-2}$. The symbols are our PIMC results; the solid line
reports the semiclassical result of the PQSCHA~\cite{cftv00}. The
dashed line is a guide for the eye (see text).
\label{figone}}
\end{figure}
The quantum dynamics of this system is ruled by the competition
between the Coulomb interaction of Cooper pairs, described by the
kinetic term, and the Josephson coupling
represented by the cosine term. The quantum fluctuations are therefore
ruled by the {\em quantum coupling} parameter
$g=\sqrt{E_{_{\rm{C}}}/E_{_{\rm{J}}}}$, where
$E_{_{\rm{C}}}=(2e)^2/2C$ is the characteristic charging energy (for
$\eta\ll{1}$). It is also convenient to use the {\em dimensionless
temperature} $t\equiv{k_{{}_{\rm B}}T/E_{_{\rm{J}}}}$. In our model
Eq.~(\ref{e.JJA}) we did not include any explicit dissipation term, as
dissipative effects are negligible provided that the shunt resistance
$R_{_{\rm{S}}}\gg{R_{_{\rm{Q}}}}\,g^2/(2\pi{t})$, where
${R_{_{\rm{Q}}}}\equiv{h/(2e)^2}$ is the quantum resistance; for
smaller $R_{_{\rm{S}}}$ the Caldeira-Leggett term can be added to the
action~(\ref{e.JJA})~\cite{fv01,cftv00}, resulting in a decrease of
quantum fluctuations.

Fig.~\ref{figone} displays our resulting phase diagram
together with the semiclassical results valid at
low coupling~\cite{cftv00}. At high temperature, the system is in
the N state with exponentially decaying phase correlations,
$\langle\varphi_{\bf{i}}\varphi_{\bf{j}}\rangle$, and vanishing
phase stiffness. By lowering $t$, for
$g\le{g^\star}\,{\simeq}\,3.4$, the system undergoes a BKT phase
transition at $t_{_{\rm{BKT}}}(g)$ to a SC state with power-law
decaying phase correlations and finite stiffness. When $g$ is
small enough (semiclassical regime), the critical temperature
smoothly decreases by increasing $g$ and it is in remarkable
agreement with the predictions of the pure-quantum self-consistent
harmonic approximation (PQSCHA)~\cite{cftv00}. For larger $g$ (but
still $g<g^\star$) the semiclassical treatment becomes less
accurate and the curve $t_{{}_{\rm BKT}}(g)$ shows a steeper
reduction, but the SC-N transition still obeys the standard BKT
scaling behavior. Finally, for $g>g^\star$ a strong quantum
coupling regime with no sign of a SC phase is found. Surprisingly,
the BKT critical temperature does not scale down to zero by
increasing $g$ (i.e., $t_{{}_{\rm{BKT}}}(g^\star)\neq{0}$): by
reducing the temperature in the region
$3.2\lesssim{g}\lesssim{g^\star}$, phase coherence is first
established, as a result of the quenching of thermal fluctuations,
and then destroyed again due to a dramatic enhancement of quantum
fluctuations near $t=0$. This is evidenced by a reentrant behavior
of the stiffness of the system, which vanishes at low and high $t$
and it is finite at intermediate temperatures. The open symbols in
Fig.~\ref{figone} mark the transition between the finite and zero
stiffness region when $t$ is lowered.

These results are obtained using PIMC simulations on $L\times L$
lattices (up to $L=96$) with periodic boundary conditions.
Thermodynamic averages are obtained by MC sampling of the partition
function after discretization of the Euclidean time
$u\in[0,\beta\hbar]$ in $P$ slices $\frac{\hbar\beta}{P}$, where $P$
is the Trotter number, using the standard Metropolis algorithm. The
actual sampling is made on imaginary-time Fourier transformed
variables using the same algorithm developed in
Ref.~\onlinecite{ccftv02}: thanks to the fact that the move amplitudes
are independently chosen and dynamically adjusted for each Fourier
component, this procedure ensures to efficiently reproduce the strong
quantum fluctuations of the paths in the region of high quantum
coupling $g$. Indeed, test simulations with the standard PIMC
algorithm showed serious problems of ergodicity, though eventually
giving the same results. The autocorrelation times has been reduced by
an over-relaxation algorithm~\cite{over} over the zero-frequency mode.

\begin{figure}[t]
\includegraphics[width=85mm]{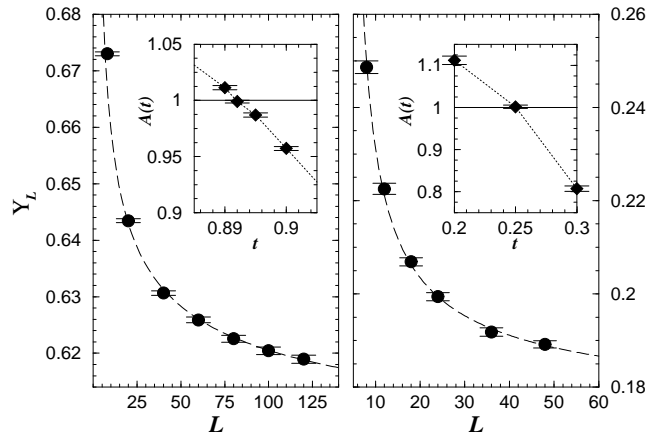}
\caption{
Size scaling of the helicity modulus $\Upsilon_L$ at the
transition temperature. Symbols are PIMC data and the dashed-lines
are the one-parameter fit with Eq.~\eqref{e.YL}. Left panel: $g=0$
and $t=0.892$ [$L_0=0.456(6)$]; right panel: $g=3.4$ and $t=0.25$
[$L_0=3.32(3)$]. The insets show $A(t)$ for different
temperatures, using the two-parameter fit (see text).
\label{figtwo}}
\end{figure}

\begin{figure}[t]
\includegraphics[width=80mm]{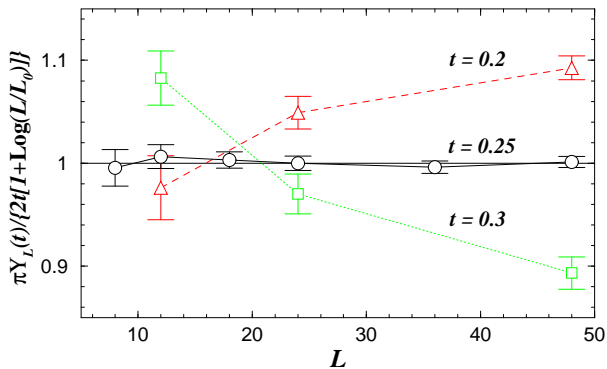}
\caption{
 Helicity modulus $\Upsilon_L$ divided by the best fit with the
 expression~(\ref{e.YL}) for $g\,{=}\,3.4$ and different
 temperatures: $\triangle$, $\bigcirc$, and $\square$ correspond to
 $t=0.2$, $0.25$, $0.3$, respectively.
\label{figthree}}
\end{figure}

A very sensitive method to determine the critical temperature is
provided by the scaling law of the helicity modulus $\Upsilon$, a
quantity proportional to the phase stiffness. $\Upsilon$ measures
the response of the system to the application of a twist $k_0$
to the boundary conditions along
a fixed direction,
\begin{equation}
 \Upsilon = \frac{1}{E_{_{\rm J}}}
\left(\frac{\partial^2 F}{\partial k_0^2}\right)_{k_0=0}~,
\label{e.Y}
\end{equation}
where $F$ is the free energy. By derivation of the discretized
path-integral expression of the partition function, the PIMC
estimator for $\Upsilon$ is easily obtained, in analogy to that of
Ref.~\cite{RJ96}. Kosterlitz's renormalization group equations
provide the critical scaling law for the finite-size
helicity modulus $\Upsilon_L$:
\begin{equation}
 \frac{\Upsilon_L(t_{{}_{\rm BKT}})}{t_{{}_{\rm BKT}}} =
 \frac{2}{\pi}\left(1+\frac{1}{2\log(L/L_0)}\right)~,
\label{e.YL}
\end{equation}
where $L_0$ is a non-universal constant. Following
Ref.~\cite{harada}, the critical temperature can be found by
fitting $\Upsilon_L(t)/t$ vs $L$ for several temperatures
according to Eq.~\eqref{e.YL} with a further multiplicative
fitting parameter $A(t)$. In this way, the critical point can be
determined by searching the temperature such that $A(t_{{}_{\rm
BKT}})\,{=}\,1$, as illustrated in Fig~\ref{figtwo}. This is the
technique we used to get the filled symbols in Fig.~\ref{figone}.
Using this procedure the critical temperature can be determined
with excellent precision. For instance in the classical case we
get $t_{{}_{\rm BKT}}(g{=}0)=0.892(2)$, in very good agreement
with the most accurate results from classical
simulations~\cite{gupta}. Also in the regime of strong quantum
coupling, $g=3.4$, the PIMC data for
$\Upsilon_L(t_{_{\rm{BKT}}}\,{=}\,0.25)$ are very well fitted by
Eq.~\eqref{e.YL}, as shown in Fig.~\ref{figthree}. Moreover, this
figure points out the sensitivity of this method to identify
$t_{_{\rm{BKT}}}$: at temperature higher (lower) than the critical
one the helicity modulus decreases (increases) much faster with
$L$ than $\Upsilon_L(t_{_{\rm{BKT}}})$. At higher values of the
quantum coupling, $g>g^\star$, the helicity modulus scales to zero
with $L\to\infty$ and $P\to\infty$ at any temperature.

\begin{figure}[t]
\includegraphics[width=80mm]{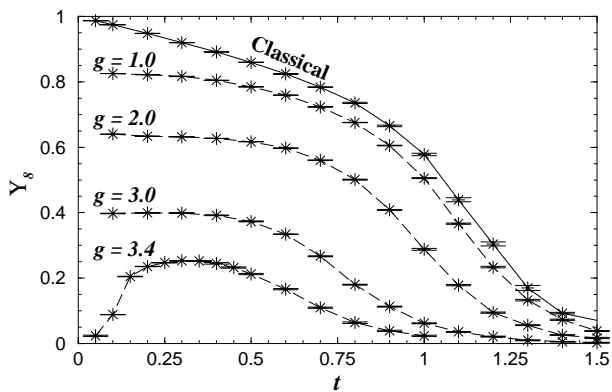}
\caption{
 Temperature behavior of the helicity modulus $\Upsilon_8(t)$ on a
 $8\,{\times}\,8$ lattice, for different values of $g$. The data
 are results from the Trotter extrapolation.
\label{figfour}}
\end{figure}

\begin{figure}[t]
\includegraphics[width=80mm]{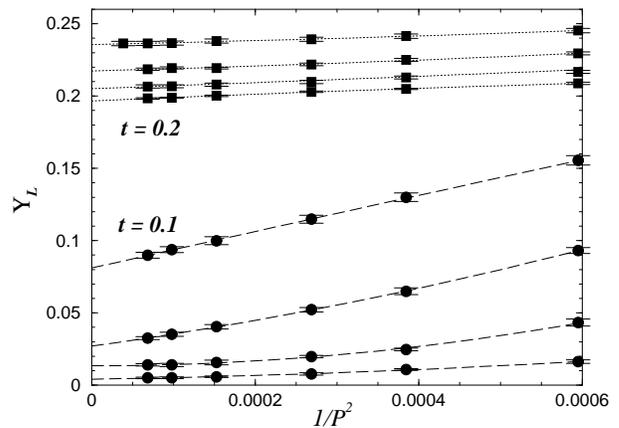}
\caption{
 Trotter-number extrapolation of $\Upsilon_L$ for $g=3.4$. Two
 series of data for $t=0.1$ ({\Large $\bullet$}) and $0.2$
 ($\blacksquare$) are reported, for four different lattice sizes:
 from the top to the bottom $L=8,10,12,14$. The lines are weighted
 quadratic fits.
\label{figfive}}
\end{figure}

\begin{figure}[t]
\includegraphics[width=80mm]{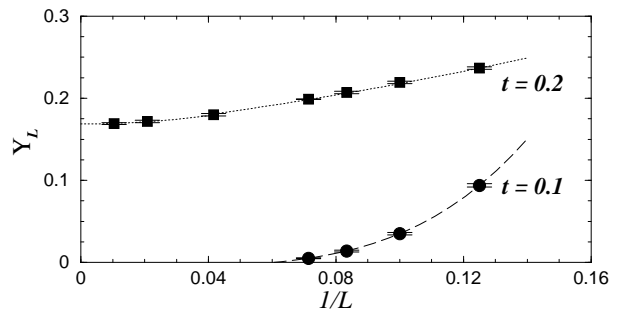}
\caption{
Finite-size scaling of the helicity modulus $\Upsilon_L$ for
$g=3.4$ at fixed $P=101$. The lines are guides for the eye.
\label{figsix}}
\end{figure}

At variance with the standard BKT theory, in the regime of strong
quantum fluctuations we find a range of coupling values,
$3.2\lesssim{g}\lesssim{g^\star}$, in which the helicity modulus
displays a non-monotonic temperature behavior. In Fig.~\ref{figfour},
$\Upsilon_L(t)$ is plotted for different values of $g$ on the
$8\times8$ cluster: up to $g=3.0$ it shows a monotonic behavior
similar to the classical case, where thermal fluctuations drive the
suppression of the phase stiffness. In contrast, for $g\,{=}\,3.4$,
the helicity modulus is suppressed at low temperature, then it
increases up to $t\sim0.2$; for further increasing temperature it
recovers the classical-like behavior and a standard BKT transition can
still be located at $t\sim0.25$ (Figs.~\ref{figtwo}
and~\ref{figthree}). A reentrance of the phase stiffness was found for
a related model in Ref.~\cite{RJ96}, but the authors concluded that
the low temperature drop of the helicity modulus was probably due to
having a finite Trotter number.  In order to ascertain this point, we
have performed systematic extrapolations in the Trotter number and in
the lattice size, as illustrated in Figs.~\ref{figfive}
and~\ref{figsix} for $g=3.4$. We have found no sign of anomalies in
the finite-$P$ behavior: the extrapolations in the Trotter number
appear to be well-behaved and already in the expected asymptotic
regime $O(1/P^2)$~\cite{trotta}, for $P\gtrsim 60$
(Fig.~\ref{figfive}). The extrapolation to infinite
lattice-size shown in Fig.~\ref{figsix} clearly indicates that
$\Upsilon_L$ scales to zero at $t=0.1$, while it remains finite and
{\em sizeable} at $t=0.2$. Hence, the outcome of our analysis is
opposite to that of Ref.~\cite{RJ96}, i.e., we conclude that the
reentrant behavior of the helicity modulus is a genuine effect present
in our model, rather than a finite-Trotter or finite-size artifact.

\begin{figure}[t]
\includegraphics[width=85mm]{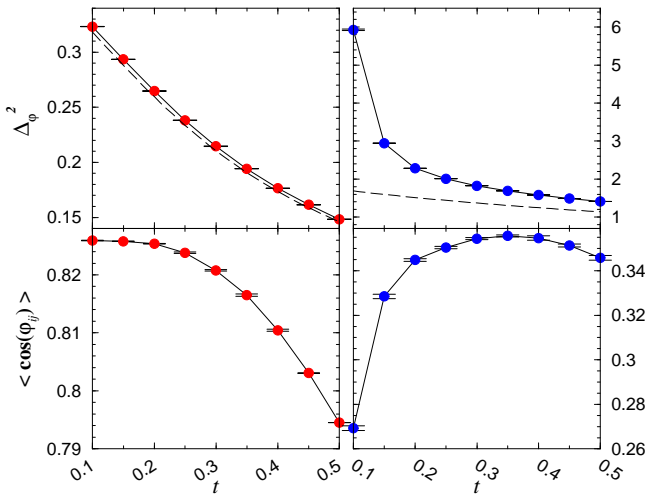}
\caption{
Top panels: $\Delta_\varphi^2$ vs $t$; bottom panels:
$\langle\cos\,\varphi_{\bm{ij}}(u)\rangle$ vs $t$. The quantum
coupling is $g=1.0$ in the left panels and $g=3.4$ in the right
panels. The circles are PIMC data and the dashed lines are PQSCHA
results.
\label{figseven}}
\end{figure}

In order to understand the physical reasons of the reentrance
observed in the phase stiffness, we have studied the following two
quantities:
\begin{eqnarray}
  && \big\langle\,\cos\,\varphi_{\bm{ij}}(u)\,\big\rangle~,
\label{e.cos}
\\
  \Delta_\varphi^2&=& \big\langle\,
 (\varphi_{\bm{ij}}(u)-\bar{\varphi}_{\bm{ij}})^2\,\big\rangle~,
\label{e.D}
\end{eqnarray}
with $\bar{\varphi}_{\bm{ij}}=(\hbar\beta)^{-1}
\int_0^{\hbar\beta}\!du~\varphi_{\bm{ij}}(u)~$
and $\bm{ij}$ nearest-neighbor sites. The first quantity is a
measure of the total (thermal and quantum) short-range
fluctuations of the Josephson phase: in particular, it has a
maximum were the overall fluctuation effect is weakest. The second
quantity represents instead the ``pure quantum'' spread of the
phase difference between two neighboring islands and has been
recently studied in the single junction problem~\cite{hz02}; more
precisely, $\Delta_\varphi^2$ measures the fluctuations around the
``static'' value (i.e., the zero-frequency component of the
Euclidean path), it is maximum at $t\,{=}\,0$ and tends to zero
in the classical limit (i.e., $g/t\to{0}$).

The quantities~(\ref{e.cos}) and~(\ref{e.D}) on a $8\,{\times}\,8$
lattice are compared in Fig.~\ref{figseven} for two values of the
quantum coupling, in the semiclassical ($g=1.0$) and in the
extreme quantum ($g=3.4$) regime. In the first case
$\langle\cos\varphi_{\bm{ij}}(u)\rangle$ decreases monotonically
by increasing $t$ and the pure-quantum phase spread
$\Delta_\varphi^2$ shows a semiclassical linear behavior which is
correctly described by the PQSCHA. At variance with this, at
$g=3.4$, where the reentrance of $\Upsilon(t)$ is observed,
$\langle\cos\varphi_{\bm{ij}}(u)\rangle$ displays a pronounced
maximum at finite temperature. Besides the qualitative agreement
with the mean-field prediction of Ref.~\cite{fms84}, we find a
much stronger enhancement of the maximum above the $t\,{=}\,0$
value. This remarkable finite-$t$ effect can be explained by
looking at $\Delta_\varphi^2(g{=}3.4)$ (Fig.~\ref{figseven},
top-right panel). Its value is an order of magnitude
higher than the one in the semiclassical approximation and,
notably, it is strongly suppressed by temperature in a
qualitatively different way from $\Delta_\varphi^2(g{=}1.0)$:
the pure-quantum contribution to the phase fluctuations
measured by $\Delta_\varphi^2$ decreases much faster than the
linearly rising classical (thermal) one. Thus the interplay
between strong quantum coupling and temperature turns out in a
finite-$t$ minimum of the total fluctuations of the Josephson
phase. This single-junction effect in a definite interval of the
quantum coupling ($3.2\lesssim{g}\lesssim3.4$) is so effective to
drive the reentrance of the phase stiffness.

As for this low-temperature transition, the open symbols in
Fig.~\ref{figone} represent the approximate location of the points
$(t,g)$ where $\Upsilon(t)$ becomes zero within the error bars: in
their neighborhood we did not find any BKT-like scaling law. This
fact opens two possible interpretations: ($i$) the transition does
not belong to the $XY$ universality class; ($ii$) it does, and in
this case the control parameter is not the (renormalized)
temperature, but a more involute function of both $t$ and $g$.
Further investigations are needed to answer this question.

In summary, we have studied a model for a JJA in the quantum
fluctuation dominated regime. The BKT phase transition has been
followed increasing the quantum coupling $g$ up to a critical
value $g^\star\,{\sim}\,3.4$ where $t_{{}_{\rm BKT}}\sim0.25$;
above $g^\star$ no traces of BKT critical behavior have been
observed. Remarkably, in the regime of strong quantum coupling
($3.2\lesssim{g}\lesssim3.4$) phase coherence is established only
in a finite range of temperatures, disappearing at higher $T$,
with a genuine BKT transition to the normal state, and at lower
$T$, due to a nonlinear quantum mechanical mechanism.

We acknowledge discussions with
G.~Falci, R.~Fazio, M.~M\"user, T.~Roscilde, and U.~Weiss. We
thank H.~Baur and J.~Wernz for assistance in using the MOSIX
cluster in Stuttgart. L.C. was supported by NSF under Grant No.
DMR02-11166. This work was supported by the MIUR-COFIN2002
program.



\begin{thebibliography}{199}
\bibitem{eta}
 The limits $\eta\ll{1}$ and $\eta\gg{1}$ are more appropriate for
 the JJA and the granular films, respectively.
\end{thebibliography}
\end{document}